\documentclass[twocolumn,pra,aps,superscriptaddress,english]{revtex4-1}
\usepackage{amsmath}
\usepackage{amssymb}
\usepackage{amsfonts}
\usepackage{amssymb}
\usepackage[pdftex]{graphicx}
\usepackage{subfigure} 
\usepackage[english]{babel}
\usepackage{braket}
\usepackage{color}
\usepackage[colorlinks,linkcolor=blue,anchorcolor=blue,citecolor=blue,urlcolor=blue]{hyperref}
\usepackage{babel}
\makeatother

\begin{document}

\renewcommand{\figurename}{FIG.}
\title{Efficient quantum microwave-to-optical conversion using electro-optic nanophotonic coupled-resonators}

\author{Mohammad Soltani}
\email{mo.soltani@raytheon.com}

\affiliation{Raytheon BBN Technologies, Cambridge, MA 02138, USA}

\author{Mian Zhang}
\email{mianzhang@seas.harvard.edu}
\affiliation{John A. Paulson School of Engineering and Applied Science, Harvard University, 29 Oxford Street, Cambridge, MA, 02138, USA}

\author{Colm Ryan}
\affiliation{Raytheon BBN Technologies, Cambridge, MA 02138, USA}

\author{Guilhem J. Ribeill}
\affiliation{Raytheon BBN Technologies, Cambridge, MA 02138, USA}

\author{Cheng Wang}
\affiliation{John A. Paulson School of Engineering and Applied Science, Harvard University, 29 Oxford Street, Cambridge, MA, 02138, USA}

\author{Marko Loncar}
%\email{loncar@seas.harvard.edu}
\affiliation{John A. Paulson School of Engineering and Applied Science, Harvard University, 29 Oxford Street, Cambridge, MA, 02138, USA}
%\selectlanguage{}%

\begin{abstract}
We propose a low noise, triply-resonant, electro-optic (EO) scheme for quantum microwave-to-optical conversion based on coupled nanophotonics resonators integrated with a superconducting qubit. Our optical system features a split resonance - a doublet - with a tunable frequency splitting that matches the microwave resonance frequency of the superconducting qubit. This is in contrast to conventional approaches where large optical resonators with free-spectral range comparable to the qubit microwave frequency are used. In our system, EO mixing between the optical pump coupled into the low frequency doublet mode and a resonance microwave photon results in an up-converted optical photon on resonance with high frequency doublet mode. Importantly, the down-conversion process, which is the source of noise, is suppressed in our scheme as the coupled-resonator system does not support modes at that frequency. Our device has at least an order of magnitude smaller footprint than the conventional devices, resulting in large overlap between optical and microwave fields and large photon conversion rate ($g/2\pi$) in the range of $\sim$5-15 kHz. Owing to large $g$ factor and doubly-resonant nature of our device, microwave-to-optical frequency conversion can be achieved with optical pump powers in the range of tens of microwatts, even with moderate values for optical $\it{Q}$ ($\sim 10^6$) and microwave $Q$  ($ \sim10^4$). The performance metrics of our device, with substantial improvement over the previous EO-based approaches, promise a scalable quantum microwave-to-optical conversion and networking of superconducting processors via optical fiber communication.  

\end{abstract}
\maketitle

\section{Introduction}

Quantum frequency conversion between superconducting (SC) microwave qubits and telecom optical photons is critical for long distance communication of networked SC quantum processors. While SC qubits operate at cryogenic temperatures to sustain their quantum coherence, converting them to the optical domain enables transferring the quantum states to room temperature and over long distances. For such a conversion process, several schemes have been investigated, including optomechanics \cite{wang2012using,andrews2014,fang2016}, magnons \cite{zhang2014strongly}, piezomechanics \cite{vainsencher2016bi,zou2016cavity}, and Pockels electro-optics (EO) \cite{tsang2010cavity,tsang2011cavity,savchenkov2010single,rueda2016efficient}.

EO conversion approach is particularly attractive since it is broadband, low noise, mechanically and thermaly stable (i.e. does not rely on free-standing structures), scalable (large scale integration of EO devices with superconducting circuits is possible), and tunable (e.g. using bias voltages). The EO effect, a $2^{nd}$ order optical nonlinearity of the material, mixes the microwave signal and the pump laser fields, thereby producing an optical frequency sideband next to the pump laser frequency  \cite{tsang2011cavity,savchenkov2010single,rueda2016efficient}. Owing to its large EO coefficient, LiNbO$_3$ (LN) is ideally suited for this task. The efficiency of the conversion process can be dramatically enhanced using an optical resonator that supports resonances at pump and sideband frequencies \cite{tsang2011cavity, savchenkov2010single,burgess2009difference,rueda2016efficient}. Whispering gallery mode (WGM) resonators fabricated by polishing LN crystal \cite{savchenkov2010single,rueda2016efficient,ilchenko2003whispering,soltani2016ultrahigh} are among the most promising candidates for this conversion process.  Still, these devices face three key limitations (see also Fig. 1(a)):
 
1-	They require that the free-spectral-range (FSR) of the optical resonator matches the microwave frequency to enable resonance-enhanced three-wave mixing with pump photons. This makes the optical resonator quite large resulting in a very small electro-optic conversion rate factor ($g/2\pi <$ 100 Hz). This can be improved using an integrated LN resonator featuring $g$ values in the range of tens of kHz, as theoretically demonstrated by \cite{javerzac2016chip},  
 
2-	They can be noisy since the mixing of the pump laser  with the microwave signal produces not only the desired up converted sideband but also the undesired down converted one. To mitigate this, previous approaches \cite{rueda2016efficient} have utilizied off-resonance pumping (see the bottom panel of Fig. 1(a)), albeit at the expense of reduced conversion efficiency.

3-	They require perfect matching between microwave resonance and the spacing between the optical resonances as there is no frequency tunability to arbitrarily space the optical resonances.

In this paper, we propose an approach that overcomes the above mentioned hurdles. Our device is based on integrated coupled-resonator system that supports a resonance doublet with a frequency splitting that matches the resonant frequency of the microwave photon, and reduces both microwave and optical mode volumes by at least an order of magnitude. We show that when implemented in an LN nanophotonic platform, our device can feature $g/2\pi$ in the $\sim$ 10 kHz range.  We emphasize that our device does not require optical FSR to SC microwave qubit frequency matching which typically results in large devices and low $g$, and consequently require off-resonance pumping to suppress unwanted down conversion process (Fig. 1(a)).  In contrast, our approach (Fig. 1(b)) is based on fully resonant process where the pump and up-converted photons are resonant with lower and higher frequency of the doublet, respectively. Importantly, the down converted photon is far detuned from any resonance of the system, and therefore this process is strongly suppressed (Fig. 1(b)). Finally, we show that optical resonance doublet frequency splitting can be controlled and tuned by applying a bias voltage which allows for perfect matching with the microwave resonance frequency.

The paper is organized as follows: Section II discusses the coupled-resonator EO device. Section III shows the numerical simulations of the $g$ factor for this device using LN integrated photonic resonators. Section IV discusses the quantum mechanical analysis of microwave-to-optical conversion with this proposed device and finds optimal condition for maximum conversion process as well as the relations between the optical and the microwave resonator parameters to the optical pump power. Section V has a discussion on other considerations relevant to this platform, and we conclude in section VI.

\section{THE PROPOSED MICROWAVE TO OPTICAL CONVERSION DEVICE\label{sec:SPATheory}}
Figure 1(b) shows the general structure of our proposed device which consists of two identical optical ring EO resonators coupled to each other by a coupling factor $\mu$. In this coupled system,  $\mu$ can be made tunable as discussed later. This coupled-resonator supports symmetric and anti-symmetric supermodes with resonance frequencies of $\omega$ ̅=$\omega_0\pm\mu$, where $\omega_0$ is the resonance frequency of the individual resonators when uncoupled. These eigenfrequencies can be found starting from the coupled-mode equations, and assuming mode amplitudes of $a_1$ and $a_2$ for each resonator:
\begin{equation}   %LEONARDO LEONARDO LEONARDO  LEONARDO  LEONARDO LEONARDO LEONARDO  LEONARDO
\frac{da_1}{dt}=-i\omega_0 a_1 + i\mu a_2
\end{equation}
\begin{equation}   %LEONARDO LEONARDO LEONARDO  LEONARDO LEONARDO LEONARDO LEONARDO  LEONARDO
\frac{da_2}{dt}=-i\omega_0 a_2 + i\mu a_1
\end{equation}

\begin{figure}
	\includegraphics[width=8cm]{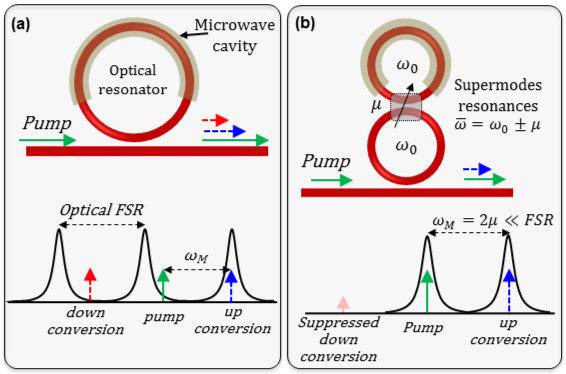}\caption{\label{fig:Schematic} Electro-optic based microwave-to-optical quantum conversion. (a) The conventional scheme wherein an electro-optic ring resonator with FSR close to but slightly different than the microwave resonance ($\omega_M$) of a SC qubit is integrated with the microwave cavity electrode for the electro-optic conversion. The pump laser slightly blue detuned from one of the resonances results in up conversion on resonance with the cavity mode, while down conversion is off-resonance, and thereby suppressed. (b) Our proposed scheme based on strongly coupled resonators features frequency doublets, corresponding to  symmetric and anti-symmetric super modes, with splitting equal to the microwave resonance. The coupling between the resonators ($\mu$) can be made tunable. In this scheme the up conversion is strongly enhanced since both pump and the up converted photon are resonant with the doublet. The down conversion is strongly suppressed since it is far detuned from the resonance of the system.}
\end{figure}

From the sum and the difference of the above two equations we find the following equations for the normal modes of the coupled resonators:
\begin{equation}   %LEONARDO LEONARDO LEONARDO  LEONARDO LEONARDO LEONARDO LEONARDO  LEONARDO
\frac{d(a_1\pm a_2)}{dt}=-i(\omega_0 \mp \mu ) (a_1 \pm a_2)
\end{equation} where $a_1+a_2$ and $a_1-a_2$ are the symmetric ($a_s$) and antisymmetric ($a_{as}$) mode of the coupled resonator with resonance frequencies $\omega_s=\omega_0-\mu$ and $\omega_{as}=\omega_0 + \mu$, respectively.

The three-wave electro-optic mixing process occurs between the super-modes of the coupled resonator system and the microwave photon mode corresponding to the SC qubit. As illustrated in Fig. 1(b), the pump laser is on resonance with the symmetric supermode, as it has the lower frequency, while the up-converted photon is resonant with the anti-symmetric supermode. The frequency splitting between the supermodes is matched to the frequency of the microwave photon ($2\mu = \omega_M$). Therefore, the $FSR$ of each resonator can be much larger than the frequency splitting ($FSR >> 2\mu$), and hence, smaller resonators with reduced mode volume can be used resulting in larger $g$. This is in contrast with conventional approaches shown in Fig. 1(a) that require a large resonator with $FSR$ approximately equal to the microwave photon frequency ($FSR \sim \omega_M$). 

The conversion rate factor ($g$) between the microwave photon and the up converted optical photon can be obtained using coupled-mode theory, by including the EO effect as a perturbative term into Maxwell equations. This results in the following rate equation
\begin{equation}   %LEONARDO LEONARDO LEONARDO  LEONARDO LEONARDO LEONARDO LEONARDO  LEONARDO
\frac{da_{as}}{dt}=-iga_sa_M
\end{equation} where $a_M$ is the mode amplitude of the microwave cavity, and $g$ can be expressed as
\begin{equation}
g = \frac{\epsilon_0\omega_0}{4U_{opt}} \sqrt{\frac{\hbar\omega_M}{U_M}}\int \bar{E}^*_{as}[\Delta\eta\bar{E}^*_{M}]E_sdv\label{eq:g}  
\end{equation}  In the above expression, the integral is calculated over the volume of the LN resonator, $\bar{E}_s$ and $\bar{E}_{as}$ are the electric fields of the symmetric and anti-symmetric modes of the coupled-resonator, $\bar{E}_M$ is the field of the microwave cavity, and $\Delta\eta$ is the electro-optic tensor of the optical resonator material. $U_M = CV^2/2$ is the microwave resonator energy, where $C$ and $V$ are the capacitance and the voltage of the microwave electrode, and $U_{opt}$ is the optical energy of the symmetric and anti-symmetric modes that have closely similar values.  Equation (5) shows that non-zero $g$ can be achieved only if microwave field is not symmetric. This can be accomplished by having microwave field interact with only one optical resonator (e.g. the upper one as shown in Fig. 1(b)) or with both resonators using a push-pull configuration. In the latter, different voltage polarity is used for each resonator to provide a non-zero electro-optic mixing. 

In order to achieve tripple-resonance condition between optical and microwave fields, it is essential to allow for dynamical tunability of the coupling factor $\mu$. Such a tuning knob would allow for any fabrication-induced changes to the resonant frequencies to be compensated for. To achieve this, we propose the configuration shown in Fig. 2(a): the resonators are coupled at two points with 100$\%$ coupling, and each resonator has a built-in phase-shifter for the tuning purpose. Using a transfer matrix analysis the symmetric and anti-symmetric eigen-resonances of this coupled resonator can be found as:
\begin{equation}   %LEONARDO LEONARDO LEONARDO  LEONARDO LEONARDO LEONARDO LEONARDO  LEONARDO
\omega_{s,as}=\omega_0 + \frac{n_gFSR}{n_{eff}}(\pi\mp\phi)
\end{equation}  where $n_{eff}$, $n_g$, and $FSR=c/(n_gl_R)$ are the effective index, group index and the free-spectral-range of the individual resonators, respectively, and $c$ and $l_R$ are the speed of light and the resonator length, respectively. The phase tuning can be achieved using the same EO effect by applying DC voltage to the electrodes of the phase shifter, without any power dissipation. Our recent work shows that a nanophotonic LN modulator similar to the ones considered here, has $V_{\pi}\cdot L \approx$ 2 V.cm. That is 2 V applied across an electrode length of $L=$ 1 cm is sufficient to provide a $\pi$ phase shift \cite{wang2017nanophotonic}. Using this number, we evaluate the resonance splitting achiavable with DC bias applied to the phase shifter assuming the $FSR = 100$ GHz for each resonator (Fig. 3). As seen from this figure, a large resonance splitting tuning range is achievable with a reasonable bias voltage. 
\begin{figure}
	\includegraphics[width=8cm]{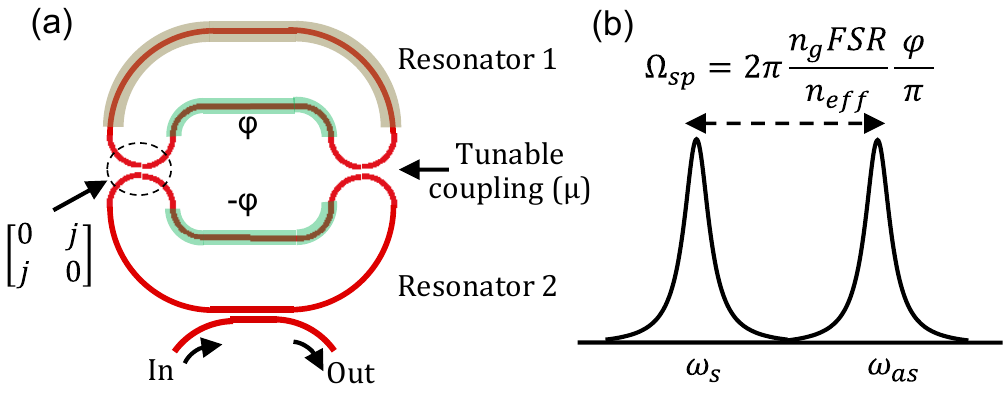}\caption{\label{fig:2}(a) Schematic of a coupled-resonator with a tunable coupling strength that can be used to dynamically control the resonance splitting. The resonators are coupled at two points using directional couplers with 100$\%$ coupling (with the coupling matrix as shown in the figure), and each has an integrated phase-shifter that imparts the phase with the opposite sign. (b) The resonance splitting can expand or contract by controlling the phase $\phi$.}
\end{figure}

\begin{figure}
	\includegraphics[width=6.5cm]{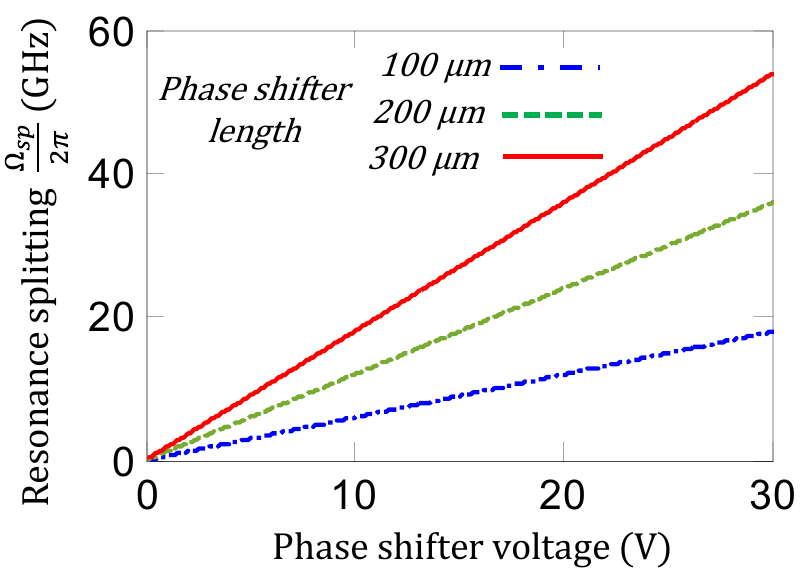}\caption{\label{fig:3}Calculated resonance splitting vs.  electrostatic voltage applied to the phase-shifter for the LN coupled-resonator with the structure shown in Fig. 2(a) and for three phase-shifter lengths of 100, 200, and 300 microns as specified in the figure. For this calculation we assume a phase shift of $\pi$ is obtained with a 2 V.cm, and use Eq. 6. We also assume the resonator has an $FSR$ of 100 GHz, $n_g = 2.39$, $n_{eff} = 2$.}
\end{figure}

\section{NUMERICAL ANALYSIS OF $g$ FACTOR\label{sec:SPATheory}}

We consider a Z-cut LN as the photonic resonator material, and two possible configurations for placing the microwave electrodes with respect to the optical resonator as illustrated in Figs. 4(a) and 4(b). The microwave electrodes are interacting only with one of the optical resonators in the coupled-resonator schemes discussed here. In both cases, optical modes are in telecom wavelength range ($\sim$ 1550 nm). The parallel electrode configuration shown in Fig. 4(a) results in larger $g$, due to stronger vertical electric field, but is somewhat more challenging to fabricate. The SC qubit is made on the silicon layer at some distance from the SiO$_2$ layer to avoid the microwave loss of SiO$_2$ \cite{OConnell2008}.  Figures 4(c) and 4(d) show the microwave field distributions for the structures in Figs. 4(a) and 4(b), respectively. It can be seen that the electric field lines inside the LN resonator are vertical, which maximizes the EO effect for the Z-cut LN, and are nearly uniform across the optical mode profile (Fig. 4(e)). This uniformity of the microwave field can be used to simplify Eq. (5) to:
\begin{equation}
g \approx n^2_e r_{33}\omega_0 \sqrt{\frac{\hbar\omega_M}{U_M}}\frac{\alpha}{2}\bar{E}^*_{Mz}
\end{equation}    where $n_e$ and $r_{33}$ are the extraordinary refractive optical index (2.138) and the electro-optic coefficient (30 pm/V) of LN, respectively, $\bar{E}_{Mz}$ is the z component of the microwave electric field at the center of the LN resonator, and $\alpha$ is the ratio of the electrode length to the perimeter of an individual resonator. The factor of 1/2 takes into account the fact that microwave field interacts only with one of the optical resonators.

\begin{figure}
	\includegraphics[width=8cm]{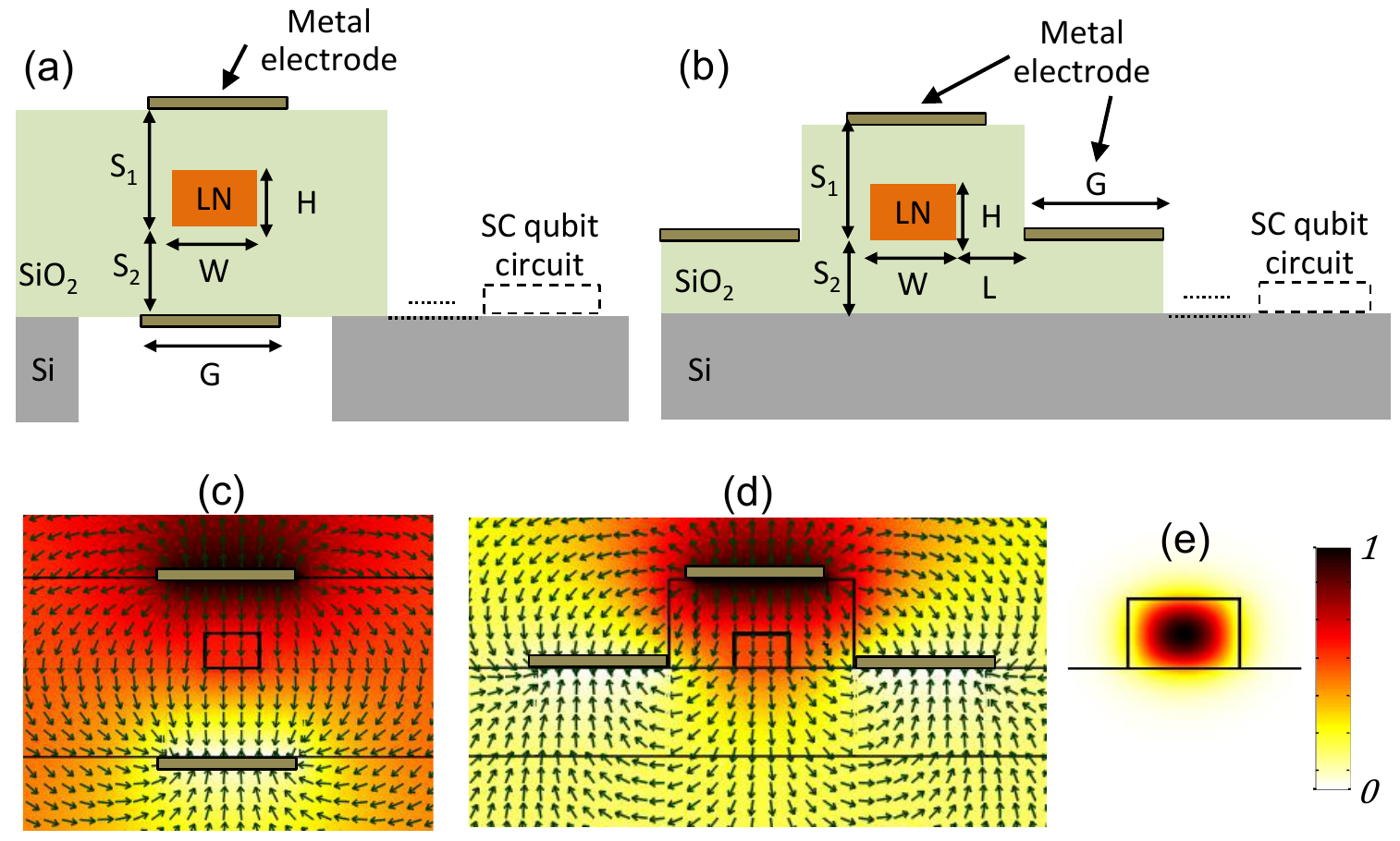}\caption{\label{fig:4}The cross section of an LN optical resonator buried in SiO$_2$ and integrated with microwave electrodes. In (a) electrodes are on top and the bottom of the LN resonator, while in (b) one of the electrodes is on top of the LN resonator, and two ground electrodes are on the sides. In both (a) and (b) the SC qubit (details not shown) is implemented on the Si substrate at some distance from the resonator. The SC qubit is capacitively connected (not shown) to the top electrode on the LN resonator via a long meander microstrip which is also and inductor. This inductor and the capacitive electrode on top of the LN resonator forms the microwave cavity. (c)-(d) show the normalized electric potential and electric field lines in the devices in (a) and (b), respectively, assuming the microwave frequency of $\omega_M=2\pi\times 6$ GHz. (e) The cross section of the normalized transverse electric optical mode profile of the LN resonator at a wavelength of 1550 nm. The simulation parameters in (c)-(e) are $W = 1.2 \mu m$, $H = 0.75 \mu m$, $S_1 = S_2 = 2 \mu m$, $G = 3 \mu m$, $L = 1.5 \mu m$. The colorbar corresponds to the figures in (c)-(e).}
\end{figure}
\begin{figure}
	\includegraphics[width=8cm]{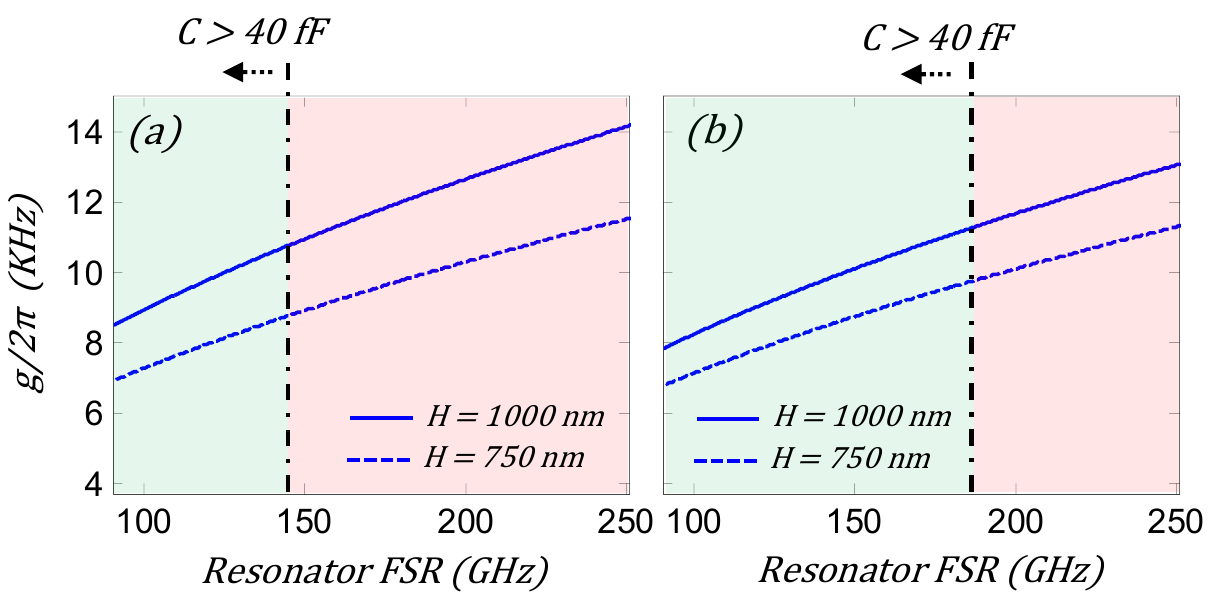}\caption{\label{fig:5}Calculated $g$ factor for the Z-cut LN coupled-resonator for different FSRs and for $\alpha = 1$. (a) The resonator has a cross section shown in Fig. 4(a) with the parameters of $W = 1.2 \mu m$, $S_1 = S_2 = 2 \mu m$, $G = 3 \mu m$, and different $H$ values as specified in the figure. (b) The resonator has a cross section shown in Fig. 4(b) with the parameters of $W = 1.2 \mu m$, $S_1 = S_2 = 2 \mu m$, $L = 1.5 \mu m$, and different $H$ values as specified in the figure. In both figures in (a) and (b) the resonator FSR values (i.e. resonator perimeters) that provide a microwave electrode capacitance $\geq 40 fF$ have been specified.}
\end{figure}
In this work we assume a microwave resonance of $\omega_M=2\pi\times 6$ GHz which is typical for SC qubits. This frequency imposes constraints on the resonance splitting required for the coupled optical resonators, as well as on the lengths of the electrodes and the transmission lines that define respectively the capacitance ($C$) and the inductance ($L$) of the microwave resonator ($\omega_M = 1/\sqrt{LC}$). With the present technology, a roughly minimum electrode capacitance of $\sim40$ fF is satisfactory to keep it large enough compared to the parasitic capacitance, which is normally $< 10$ fF, and to make a decent meander length inductance on chip.

\begin{figure}
	\includegraphics[width=8cm]{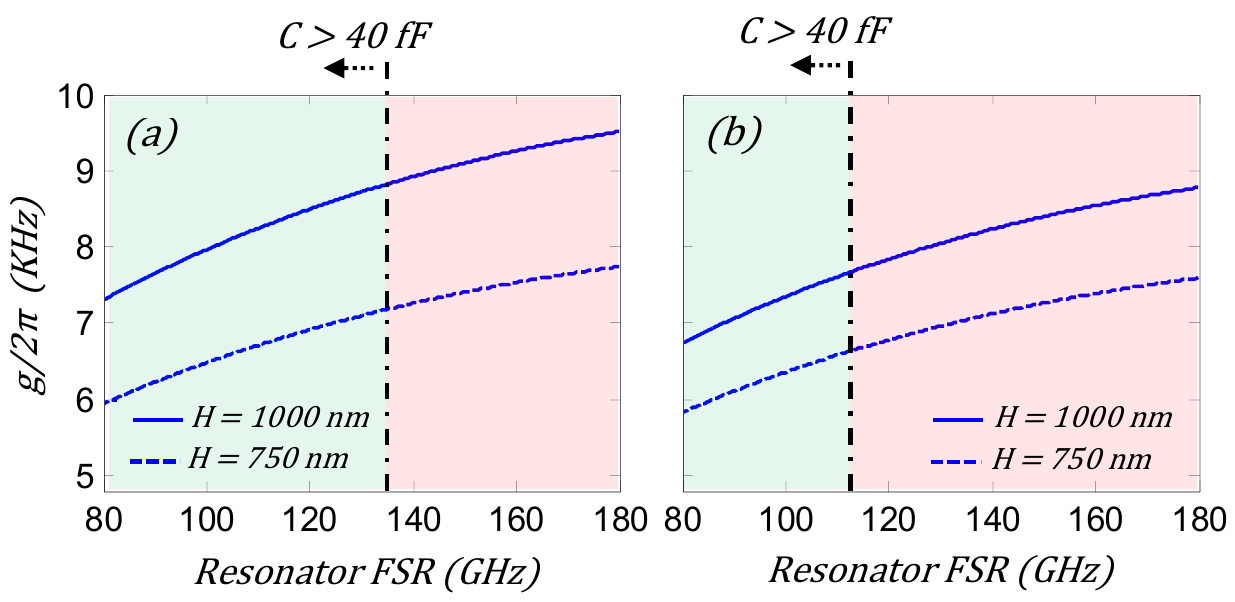}\caption{\label{fig:6}Calculated $g$ factor for the Z-cut LN coupled-resonator with tunable coupling as shown in Fig. 2(a) versus the FSR and for a phase-shifter lengths of 200$\mu m$, and $\alpha = 0.55$. (a) The resonator has a cross section shown in Fig. 4(a) with the parameters of $W = 1.2 \mu m$, $S_1 = S_2 = 2 \mu m$, $G = 3 \mu m$, and different $H$ values as specified in the figure. (b) The resonator has a cross section shown in Fig. 4(b) with the parameters of $W = 1.2 \mu m$, $S_1 = S_2 = 2 \mu m$, $L = 1.5 \mu m$, and different $H$ values as specified in the figure. In both figures in (a) and (b) the resonator FSR values (i.e. resonator perimeters) that provide a microwave electrode capacitance $> 40 fF$ have been specified.  }
\end{figure}
Figures 5(a) and 5(b) show the calculated $g$ factor for the coupled-resonator, without tunable coupling, and with microwave electrode configurations depicted in Fig. 4(a) and 4(b), respectively, for different resonator $FSR$s (i.e. resonator lengths). For these calculations we assume the microwave electrode is on one of the optical resonators and covering its entire perimeter, i.e. $\alpha = 1$ in Eq. (7). We find $g/2\pi \approx$ 10 kHz, which is quite promising given the compact resonator sizes. As expected, the $g$ values in Fig. 5(a) are larger than those in Fig. 5(b) owing to a stronger microwave electric field in the parallel electrode geometry shown in Fig. 4(a). 

Figures 6(a) and 6(b) shows the calculated $g$ for the coupled-resonator scheme shown in Fig. 2(a) that features phase-shifter section in each ring resonator, and for microwave electrode configurations depicted in Fig. 4(a) and 4(b), respectively. The $g$ values in Fig. 6 are relatively smaller than the ones in Fig. 5, since the SC microwave electrode is absent in the coupling region, thus reducing the overall interaction length between the microwave and the optical modes, resulting in $\alpha=$ 0.55. Furthermore, since $U_M = 1/2 CV^2$, then $g\sim\sqrt{1/C}\sim\sqrt{1/l}$ (see Eq. (7)), where $l$ is the electrode length. Therefore, due to $~\sqrt{1/l}$ dependence, the $g$ values in Fig. 6 are not significantly different from those in Fig. 5. Further optimization of the coupled-resonator geometry and the phase-shifter can provide simultaneously large $g$ and wide tuning range.

\section{QUANTUM MECHANICAL ANALYSIS AND OPTIMAL CONDITIONS OF THE CONVERSION PROCESS\label{sec:Quantum}}

In this section we analyze the microwave-to-optical conversion process at a single photon level, and find the optimal conditions with respect to the optical and the microwave resonator parameters as well as the optical pump power. For simplicity, we consider the geometry without the phase shifters, as illustrated in Figure 7. The energy Hamiltonian ($\hat{H}$) of the device includes the interaction of two doublet optical modes and the microwave mode in the presence of the optical and the microwave waveguides. For the analysis we consider two scenarios: 1- a closed quantum system where the waveguides are absent, and the resonators do not have any loss, and 2- an open quantum system where the resonators are coupled to the waveguides as well as a continuum reservoir of loss.

\subsection{A Closed Quantum System Analysis}
\begin{figure}
	\includegraphics[width=8cm]{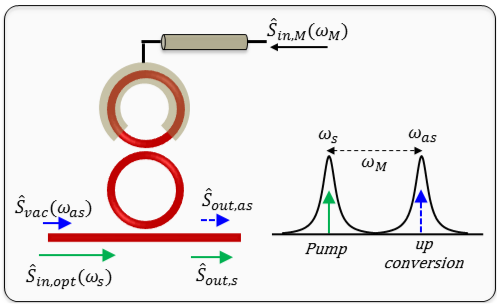}\caption{\label{fig:7}Schematic of the coupled-resonator microwave-to-optical conversion with input and out parameters specified. The optical pump at a frequency of $\omega_s$ and the vacuum mode at frequency $\omega_{as}$ are coupled in from a waveguide to the symmetric and anti-symmetric modes of the coupled-resonator, respectively.}
\end{figure}
In the absence of the waveguides and any external loss, $\hat{H}$ can be written as the sum of the non-interacting ($\hat{H_0}$) and interacting ($\hat{V}$) terms as:
\begin{equation} %LEONARDO LEONARDO LEONARDO  LEONARDO LEONARDO LEONARDO LEONARDO  LEONARDO
\hat{H} = \hat{H_0}+\hat{V}
\end{equation} where
\begin{equation} %LEONARDO LEONARDO LEONARDO  LEONARDO LEONARDO LEONARDO LEONARDO  LEONARDO
\hat{H_0} = \hbar\omega_s\hat{a}^\dagger_s \hat{a}_s + \hbar\omega_{as}\hat{a}^\dagger_{as} \hat{a}_{as}  +  \hbar\omega_M\hat{a}^\dagger_M \hat{a}_M
\end{equation} and
\begin{equation} %LEONARDO LEONARDO LEONARDO  LEONARDO LEONARDO LEONARDO LEONARDO  LEONARDO
\hat{V} = \hbar g(\hat{a}^\dagger_{as}\hat{a}_M \hat{a}_s + adjoint)
\end{equation}   In the above expressions $\omega_s$ and $\omega_{as}$ are the 
optical doublet resonance corresponding to the symmetric and anti-symmetric mode 
of the coupled-resonator, respectively, and $\omega_M$ is the resonance of the 
microwave cavity. $\hat{a}_s$, $\hat{a}_{as}$, and $\hat{a}_M$ (and their adjoints) 
are the anihiliation (creation) operators for these resonance modes, and $g$ is 
the coupling factor that mixes the microwave mode to the optical mode as 
discussed in earlier sections. By inserting the above expressions into the 
Heisenberg picture equation of motion:
\\ 
\begin{equation}
\frac{d\hat{a}_x}{dt} =\frac{i}{\hbar}[\hat{H},\hat{a}_x]   \,,\,   (x = s,as, M)
\end{equation}  
\\
we obtain the following dynamical equations:
\\
\begin{align}  
\frac{d\hat{a}_s}{dt} &=-i\omega_s\hat{a}_s - ig\hat{a}^\dagger_M\hat{a}_{as}\\
\frac{d\hat{a}_{as}}{dt} &=-i\omega_{as}\hat{a}_{as} - ig\hat{a}_M\hat{a}_{s}\\
\frac{d\hat{a}_M}{dt} &=-i\omega_M\hat{a}_M - ig\hat{a}^\dagger_s\hat{a}_{as}
\end{align}   
\\
By replacing the above operators with their slowly varying terms as 
$\hat{a}_x=\hat{A}_x  exp(-i\omega_x t)$ we have the following equations:
\\
\begin{align}
\frac{d\hat{A}_s}{dt} &=- ig\hat{A}^\dagger_M\hat{A}_{as}\\
\frac{d\hat{A}_{as}}{dt} &=- ig\hat{A}_M\hat{A}_{s} \label{eq:As}\\
\frac{d\hat{A}_M}{dt} &=- ig\hat{A}^\dagger_s\hat{A}_{as} \label{eq:AM}
\end{align}  
\\
Assuming that $\hat{A}_{s}$ carries the pump signal and in the non-depletion 
mode, and so can be treated classically, the Eqs. (16) and (17) are simplified as:
\begin{align} 
\frac{d^2\hat{A}_{as}}{dt^2} &=- g^2N_s\hat{A}_{as}\\
\frac{d^2\hat{A}_{M}}{dt^2} &=- g^2N_s\hat{A}_{M}
\end{align} 
\\  
where $N_s=<\hat{A}^\dagger_{s} \hat{A}_s>$ is the total number of pump photons 
inside the resonator. The above two differential equations for this closed 
quantum system have sinusoidal solutions with a frequency $g\sqrt{N_s}$. If we 
assume an initial value of $\hat{A}_{M0}$ and $\hat{A}_{as0}$ for the quantum 
microwave and the optical states at t = 0, the evolution of these states can be 
found as
\begin{equation}
\hat{A}_M= \hat{A}_{M0}cos(g\sqrt{N_s}t) - ie^{-i\phi_s} \hat{A}_{as0}sin(g\sqrt{N_s}t)
\end{equation}
\begin{equation}
\hat{A}_{as}= \hat{A}_{as0}cos(g\sqrt{N_s}t) - ie^{+i\phi_s} \hat{A}_{M0}sin(g\sqrt{N_s}t)
\end{equation}  
\\
where $\phi_s$ is the phase of the pump photons. From Eqs. (20) and (21) we see that complete quantum state conversion from the 
microwave to optical domain happens after a time $T=\pi/(2g\sqrt{N_s})$, where $\hat{A}_{as} 
(T)\propto ̂\hat{A}_{M0}$ and $\hat{A}_M (T)\propto \hat{A}_{as0}$. Therefore, once the microwave photon is converted to the optical 
photon, it needs to leave the coupled-resonator through 
an optical waveguide before it is converted back to the microwave domain. This is discussed in the next section. From Eq. (20) and 
(21) we see that the conversion frequency $g\sqrt{N_s}$ is dependent on the 
geometric parameters of the optical and microwave resonators and the 
electro-optic coefficient, all summarized in $g$ factor, and number of optical pump photons 
inside the resonator $(N_s)$.
\subsection{An Open Quantum System Analysis}  

Next, we consider an open quantum system where both optical and microwave resonators are lossy and interact with external optical and microwave waveguides respectively, as shown in Fig. 7. In this case, using the quantum Langevin dynamic equations and the input-output formalism \cite{tsang2011cavity,ilchenko2003whispering,walls2007quantum}, the system of equations becomes: 
\begin{equation}
\frac{d\hat{A}_s}{dt} =-\frac{\gamma_{opt}}{2}\hat{A}_s- ig\hat{A}^\dagger_M\hat{A}_{as} + \sqrt{\gamma_{ex,opt}}\hat{S}_{in,opt}
\end{equation}
\begin{equation}
\frac{d\hat{A}_{as}}{dt} =-\frac{\gamma_{opt}}{2}\hat{A}_{as}- ig\hat{A}_M\hat{A}_s + \sqrt{\gamma_{ex,opt}}\hat{S}_{vac} + \sqrt{\gamma_{i,opt}}\hat{S}'_{vac}
\end{equation}
\begin{equation}
\frac{d\hat{A}_M}{dt} =-\frac{\gamma_M}{2}\hat{A}_M- ig\hat{A}^\dagger_s\hat{A}_{as} + \sqrt{\gamma_{ex,M}}\hat{S}_{in,M} + \sqrt{\gamma_{i,M}}\hat{S}''_{vac}
\end{equation}
\begin{equation}
\hat{S}_{out,as} =\hat{S}_{vac} - \sqrt{\gamma_{ex,opt}}\hat{A}_{as}
\end{equation}  where $\gamma_{opt} = \gamma_{i,opt}+\gamma_{ex,opt}$ is optical and $\gamma_M = \gamma_{i,M}+\gamma_{ex,M}$ microwave decay rate, and the subscripts $i$ and $ex$ indicate the intrinsic and extrinsic loss rates, respectively. The later are due to coupling to the optical waveguide and the microwave transmission line. These decay rates are related to their corresponding optical and microwave quality factors as $Q_{x,opt}=\omega_{opt}/(\gamma_{x,opt})$ and $Q_{x,M}=\omega_M/(\gamma_{x,M})$. 

$\hat{S}_{in,opt}$ and $\hat{S}_{in,M}$ are the mode amplitudes of the optical pump (coupled to the optical waveguide) and the microwave signal (coupled to the microwave transmission line) which excite the symmetric mode of the coupled-resonator and the microwave cavity, respectively. They are related to their mode power as below: 
\begin{equation}
P_{in,opt} = \hbar\omega_{opt}<\hat{S}^\dagger_{in,opt}\hat{S}_{in,opt}>
\end{equation}
\begin{equation}
P_{in,M} = \hbar\omega_M<\hat{S}^\dagger_{in,M}\hat{S}_{in,M}>
\end{equation}

The anti-symmetric mode of the coupled-resonator in Eq. (23) is coupled to two Langevin vacuum noise sources: 1- the vacuum mode of the optical waveguide ($\hat{S}_{vac}$) that couples to the resonator. 2- The noise due to the intrinsic loss of the resonator coupled to the background thermal bath ($\hat{S}'_{vac}$). The microwave cavity mode in Eq. (24) is also coupled to Langevin noise ($\hat{S}''_{vac}$) due to the intrinsic cavity loss. The Langevine noise term in Eq. (22) is negligible compared to the excitation source $\hat{S}_{in,opt}$, and so is not included. Eq. (25) describes the decay of the anti-symmetric mode, which carries the up converted photon, to the optical waveguide.   

Again assuming that the pump field $\hat{A}_s$ is not depleted, and so can be treated classically, the solutions to Eq. (23) and (24) can be obtained in the frequency domain as
%\begin{equation}
%H_{{\rm int}}=i\hbar\int_{0}^{\infty}d\omega[g(\omega)e^{i\omega z_{0}/c}\hat{a}(\omega)\hat{\sigma}_{+}-{\rm h.c.}],\label{eq:H_int}
%\end{equation}

%\begin{equation}
%\begin{aligned}
%\hat{A}_M(\Omega)&=\frac{-ig\sqrt{N_s}e^{i\Phi_s} 
%	\sqrt{2\gamma_{ex,M}}}{\left(i\Omega+\gamma_{opt}\right) 
%	\left(i\Omega+\gamma_M\right)+g^2N_s} \hat{S}_{in,M}\\
% &+ \frac{(i\Omega+\gamma_M) \sqrt{2\gamma_{ex,opt}} } {(i\Omega+\gamma_{opt}) 
%	(i\Omega+\gamma_M)+g^2N_s} 
%\end{aligned}
%\end{equation}
\begin{equation}
\begin{aligned}
\hat{A}_{as}(\Omega)&=\frac{-ig\sqrt{N_s}e^{i\phi_s} 
	\sqrt{\gamma_{ex,M}}}{(i\Omega+\gamma_{opt}/2) 
	(i\Omega+\gamma_M/2)+g^2N_s} \hat{S}_{in,M}(\Omega)\\
&+ \frac{-ig\sqrt{N_s}e^{i\phi_s} 
	\sqrt{\gamma_{i,M}}}{(i\Omega+\gamma_{opt}/2) 
	(i\Omega+\gamma_M/2)+g^2N_s} \hat{S}''_{vac}(\Omega)\\
&+ \frac{(i\Omega+\gamma_M/2) \sqrt{\gamma_{ex,opt}} } {(i\Omega+\gamma_{opt}/2) 
	(i\Omega+\gamma_M/2)+g^2N_s}\hat{S}_{vac}(\Omega)\\
&+ \frac{(i\Omega+\gamma_M/2) \sqrt{\gamma_{i,opt}} } {(i\Omega+\gamma_{opt}/2) 
	(i\Omega+\gamma_M/2)+g^2N_s}\hat{S}'_{vac}(\Omega)  
\end{aligned}
\end{equation}
\begin{equation}
\begin{aligned}
\hat{A}_M(\Omega)&=\frac{(i\Omega+\gamma_{opt}/2) \sqrt{\gamma_{ex,M}} } {(i\Omega+\gamma_{opt}/2) 
	(i\Omega+\gamma_M/2)+g^2N_s}\hat{S}_{in,M}(\Omega) \\
&+ \frac{(i\Omega+\gamma_{opt}/2) \sqrt{\gamma_{i,M}} } {(i\Omega+\gamma_{opt}/2) 
	(i\Omega+\gamma_M/2)+g^2N_s} \hat{S}''_{vac}(\Omega) \\
&- \frac{ig\sqrt{N_s}e^{-i\phi_s} 
	\sqrt{\gamma_{ex,opt}}}{(i\Omega+\gamma_{opt}/2) 
	(i\Omega+\gamma_M/2)+g^2N_s} \hat{S}_{vac}(\Omega)\\
&- \frac{ig\sqrt{N_s}e^{-i\phi_s} 
	\sqrt{\gamma_{i,opt}}}{(i\Omega+\gamma_{opt}/2) 
	(i\Omega+\gamma_M/2)+g^2N_s} \hat{S}'_{vac}(\Omega)\\
\end{aligned}
\end{equation}
, and if we are exactly 
on resonance (i.e. $\Omega=0$), then the above two equations become: 
\begin{equation}
\begin{aligned}
\hat{A}_{as}&=\frac{-ig\sqrt{N_s}e^{i\phi_s} 
	\sqrt{\gamma_{ex,M}}}{\gamma_{opt}\gamma_M/4+g^2N_s} \hat{S}_{in,M}\\
&+ \frac{-ig\sqrt{N_s}e^{i\phi_s} 
	\sqrt{\gamma_{i,M}}}{\gamma_{opt}\gamma_M/4+g^2N_s} \hat{S}''_{vac}\\
&+ \frac{\gamma_M/2 \sqrt{\gamma_{ex,opt}} } {\gamma_{opt}\gamma_M/4+g^2N_s}\hat{S}_{vac} \\
&+ \frac{\gamma_M/2 \sqrt{\gamma_{i,opt}} } {\gamma_{opt}\gamma_M/4+g^2N_s}\hat{S}'_{vac} \\
\end{aligned}
\end{equation}
\begin{equation}
\begin{aligned}
\hat{A}_M&=\frac{\gamma_{opt}/2 \sqrt{\gamma_{ex,M}} } {\gamma_{opt} 
	\gamma_M/4+g^2N_s}\hat{S}_{in,M} \\
&+\frac{\gamma_{opt}/2 \sqrt{\gamma_{i,M}} } {\gamma_{opt} 
	\gamma_M/4+g^2N_s}\hat{S}''_{vac} \\
&- \frac{ig\sqrt{N_s}e^{-i\phi_s} 
	\sqrt{\gamma_{ex,opt}}}{\gamma_{opt} 
	\gamma_M/4+g^2N_s} \hat{S}_{vac}\\
&- \frac{ig\sqrt{N_s}e^{-i\phi_s} 
	\sqrt{\gamma_{i,opt}}}{\gamma_{opt} 
	\gamma_M/4+g^2N_s} \hat{S}'_{vac}
\end{aligned}
\end{equation}  Finally, by putting Eq. (30) into Eq. (25), we obtain the following expression for the up converted photon mode in the optical waveguide
\begin{equation}
\begin{aligned}
\hat{S}_{out,as}&=\frac{ig\sqrt{N_s}e^{i\phi_s} 
	\sqrt{\gamma_{ex,M}\gamma_{ex,opt}}}{\gamma_{opt}\gamma_M/4+g^2N_s} \hat{S}_{in,M}\\
&+ \frac{ig\sqrt{N_s}e^{i\phi_s} 
	\sqrt{\gamma_{ex,opt}}\sqrt{\gamma_{i,M}}}{\gamma_{opt}\gamma_M/4+g^2N_s} \hat{S}''_{vac}\\
&+ \frac{\gamma_M (\gamma_{i,opt}-\gamma_{ex,opt})/4+g^2N_s} {\gamma_{opt}\gamma_M/4+g^2N_s}\hat{S}_{vac}\\
&- \frac{\gamma_M/2 \sqrt{\gamma_{ex,opt}}\sqrt{\gamma_{i,opt}} } {\gamma_{opt}\gamma_M/4+g^2N_s}\hat{S}'_{vac} \\ 
\end{aligned}
\end{equation}
 
From either Eq. (30) or (32), and ignoring the vacuum terms, the maximum conversion of the microwave ($\hat{S}_{in,M}$) to optical ($\hat{S}_{out,as}$) photons occurs when $4g^2 N_s/(\gamma_{opt} \gamma_M)$ equals to 1. This expression is also knows as the cooperativity factor: 
\begin{equation}
C = 4g^2N_s/(\gamma_{opt}\gamma_M)
\end{equation}
Putting Eq. (33) when $C = 1$ into Eq. (32), we find the optimal up conversion amplitude as:
\begin{equation}
\begin{aligned}
&\hat{S}_{out,as}= ie^{i\phi_s}\sqrt{\frac{\gamma_{ex,M}\gamma_{ex,opt}}{\gamma_{M}\gamma_{opt}}}\hat{S}_{in,M} + \frac{\gamma_{i,opt}}{\gamma_{opt}}\hat{S}_{vac}\\ &-\frac{\sqrt{\gamma_{ex,opt}\gamma_{i,opt}}}{\gamma_{opt}} \hat{S}'_{vac}+ie^{i\phi_s}\sqrt{\frac{\gamma_{ex,opt}\gamma_{i,M}}{\gamma_{M}\gamma_{opt}}} \hat{S}''_{vac}   
\end{aligned} 
\end{equation} From Eq. (34) we find the net conversion factor from the microwave to the optical photon as 
\begin{equation}
\begin{aligned}
\eta_{M\rightarrow opt}&=\frac{<\hat{S}^\dagger_{out,as}\hat{S}_{out,as}>}{<\hat{S}^\dagger_{in,M}\hat{S}_{in,M}>}=\frac{\gamma_{ex,opt}\gamma_{ex,M}}{\gamma_{opt}\gamma_M}\\
&=\frac{Q_{opt}Q_M}{Q_{ex,opt}Q_{ex,M}}
\end{aligned}
\end{equation}

From Eq. (35) we see that the conversion factor can approach $\sim1$ when both the optical and the microwave resonators are in strongly overcoupled regime (i.e. $\gamma_{ex,opt}\gg \gamma_{i,opt}$ and $\gamma_{ex,M}\gg \gamma_{i,M}$ such that $\gamma_{opt}=\gamma_{ex,opt}$ and $\gamma_M=\gamma_{ex,M}$). Operation at such regime also dramatically reduces the noise terms in Eq. (34). However, the optical pump power increases when the optical resonator is strongly overcoupled as discussed later. On the other hand and at the critical coupling regime for both the optical and the microwave resonators (i.e. $\gamma_{ex,opt}=\gamma_{i,opt}$ and $\gamma_{ex,M}=\gamma_{i,M}$), the conversion factor from Eq. (35) is 0.25, meaning that out of every 4 microwave photons, only one of them is converted to the optical domain. In addition, in this regime, the effect of noise terms in Eq. (34) is considerable. Therefore, an optimal choice of the coupling regime needs to be found. The choice of the coupling regime will also influence the required optical pump power.

To understand the influence of optical pump power on the conversion process we need to solve Eq. (22). In this equation, the term $\hat{A}^\dagger_M\hat{A}_{as}$ is negligible, and so a simple solution for $\hat{A}_s$ can be found in the frequency domain as 
\begin{equation}
\hat{A}_{s}(\Omega)=\frac{\sqrt{\gamma_{ex,opt}}} {(i\Omega+\gamma_{opt}/2)} \hat{S}_{in,opt}(\Omega)
\end{equation}  Using the above equation on resonance ($\Omega=0$) and Eq. (26), the total number of pump photons inside the resonator is related to the pump power inside the waveguide as
\begin{equation}
N_s=<\hat{A}^\dagger_{s} \hat{A}_s> = \frac{4\gamma_{ex,opt}}{\gamma^2_{opt}}\frac{P_{in,opt}}{\hbar\omega_{opt}} = \frac{4Q^2_{opt}}{Q_{ex,opt}}\frac{P_{in,opt}}{\hbar\omega^2_{opt}}
\end{equation}   By putting Eq. (37) into the optimal cooperativity expression ($C = 1$) in Eq. (33) we find the following expression for the input pump power in the optical waveguide
\begin{equation}
P_{in,opt}=\hbar\omega_{opt}\frac{\gamma^3_{opt}\gamma_M}{16\gamma_{ex,opt}g^2} = \hbar\omega_M\frac{\omega^3_{opt}Q_{ex,opt}}{16Q^3_{opt}Q_Mg^2}
\end{equation}

In practical applications it is important to reduce the pump power to minimize the noise and possibly detrimental effects of scattered optical photons on the SC qubit (e.g. by breaking Cooper pairs). As seen from the above expression, increasing the optical $Q$ and $g$ has more impact than increasing the microwave $Q$ on reducing $P_{in,opt}$ . Using Eqs. (37) and (38) the scattered power can be expressed as 
\begin{equation}
\begin{aligned}
P_{scat}&=\gamma_{i,opt}\hbar\omega_{opt}N_s = \frac{4Q^2_{opt}}{Q_{ex,opt}Q_{i,opt}}P_{in,opt}\\
&=\hbar\omega_M\frac{\omega^3_{opt}}{4Q_{i,opt}Q_{opt}Q_Mg^2}
\end{aligned}
\end{equation} From Eq. (39) we see that at critical coupling $P_{scat} = P_{in,opt}$ as expected, and in the strongly overcoupled regime, $P_{scat} = 4Q_{ex,opt}/Q_{i,opt}P_{in,opt}$. However, as shown later, operation in the latter case dramatically increases the optical pump power.

Eqs. (33)-(35) and (37)-(39) are the key equations that can be used to evaluate the performance of a coupled-resonator device for efficient microwave-to-optical conversion process. We use these equations in the following to evaluate the device performance at different conditions.

\begin{figure}
	\includegraphics[width=8cm]{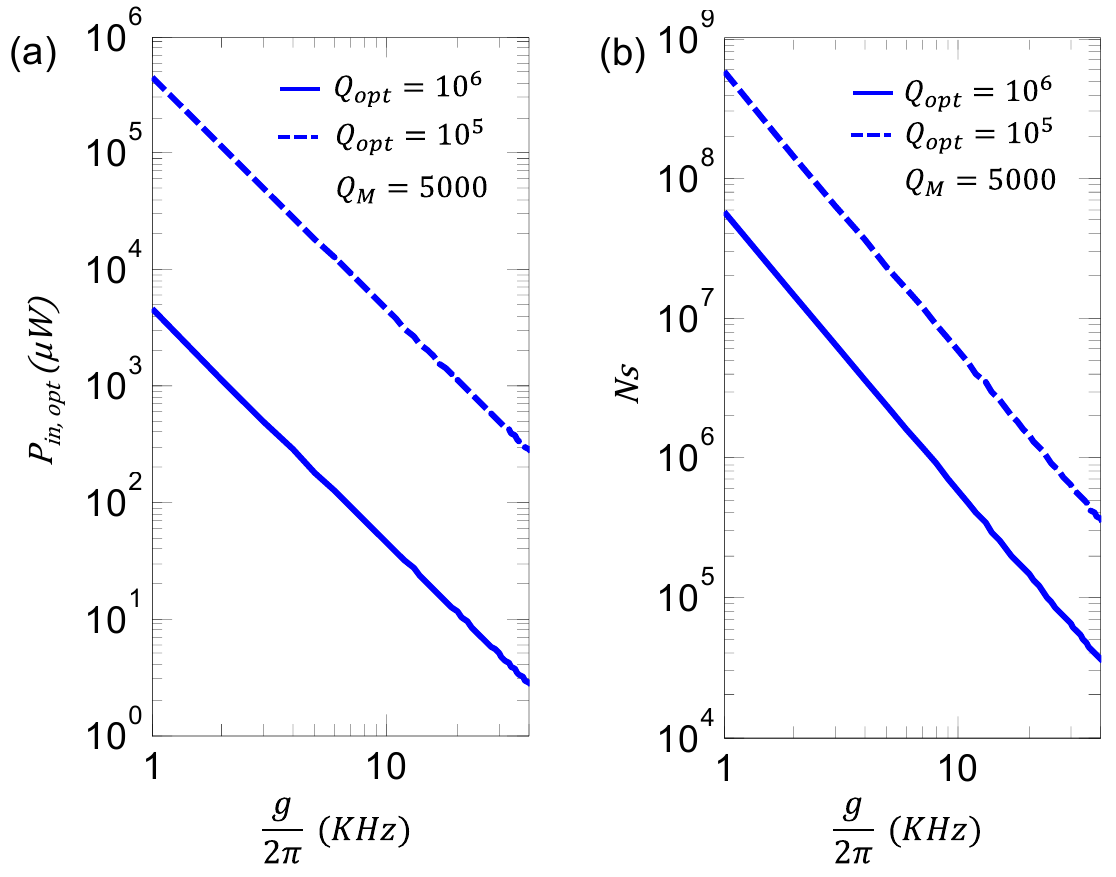}\caption{\label{fig:8}Plots of (a) the optical pump power in the waveguide (Eq.(38)) and (b) the pump photon number inside the resonator (Eq.(33)) at different $g$ rates. The solid plot corresponds to $Q_{opt}=Q_{ex,opt}/2=Q_{i,opt}/2=10^6$ and $Q_M=Q_{ex,M}/2=Q_{i,M}/2=5000$. The dashed plot corresponds to $Q_{opt}=Q_{ex,opt}/2=Q_{i,opt}/2=10^5$ and the same $Q_M$ as in (a). Both the results in (a) and (b) are for the optimal cooperativity ($C = 1$).  }
\end{figure}

Figure 8(a) shows the optical waveguide pump power at different $g$ values when operating at the optimal cooperativity condition ($C=1$). We assume both the optical and the microwave resonators operate at the critical coupling regime with quality factors within the range of the present technology as specified in the figure. From this figure we see that a pump power in tens of microwatt range is enough for the efficient conversion process when $g\approx 2\pi\times10$ kHz. Being in the critical coupling regime, the scattered photon power is equal to the optical pump power, and as shown in Fig. 8(a), it can be quite small. Furthermore, we can optimize the location of the SC qubit to minimize the impact of scattered optical photons.  Figure 8(b) shows the number of pump photons inside the resonator. However, operation in the critical coupling for both the optical and the microwave resonators can increase the effect of noise terms as seen in Eq.(34). The maximum suppression of the noise terms in Eq. (34) requires both the optical and the microwave resonators to be in strongly overcoupled regime. 

\begin{figure}
	\includegraphics[width=6.5cm]{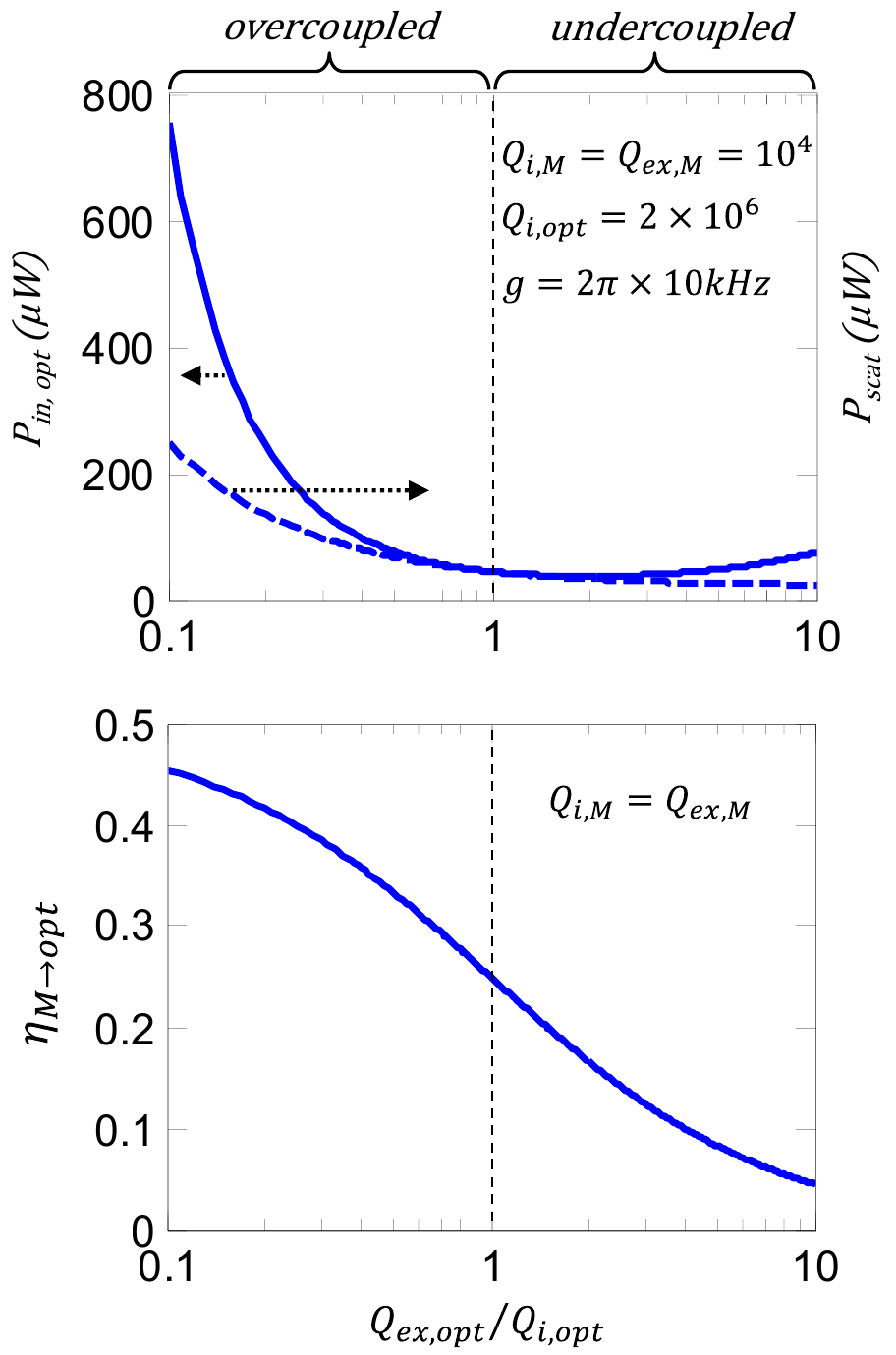}\caption{\label{fig:9}Plots of (a) the optical pump power in the waveguide (Eq.(38)), and the scattered (lost) power from the coupled-resonator (Eq.(39)), and (b) the microwave-to-optical conversion factor (Eq.(35)) for different coupling $Q$ rates between the optical resonator and the waveguide characterized by $Q_{ex,opt}$. These plots and their corresponding equations are for the optimal cooperativity (C = 1). The other device parameters are shown in the insets of the figures.}
\end{figure}

An interesting study is to investigate the optimal optical pump power, scattered power, and the conversion factor for different range of coupling of the optical resonator to the external waveguide. This is shown in Fig. 9 by varying $Q_{ex,opt}$ and fixing the other parameters. The regimes of over-coupling and under-coupling have been indicated, as well as the critical coupling case of $Q_{ex,opt}=Q_{i,opt}$ (vertical dashed line). It can be seen that operating in strongly overcoupled regime improves the conversion factor (Fig. 9(b)), though it comes at the expense of larger pump power (Fig. 9(a)), and thus larger scattered power. Therefore, a trade-off exists between increasing the conversion factor and reducing the pump power. As a design rule, we can operate in the overcoupled regime as long as the increased scattered power does not degrade the SC qubit by heating effects. 

\begin{figure}
	\includegraphics[width=6.6cm]{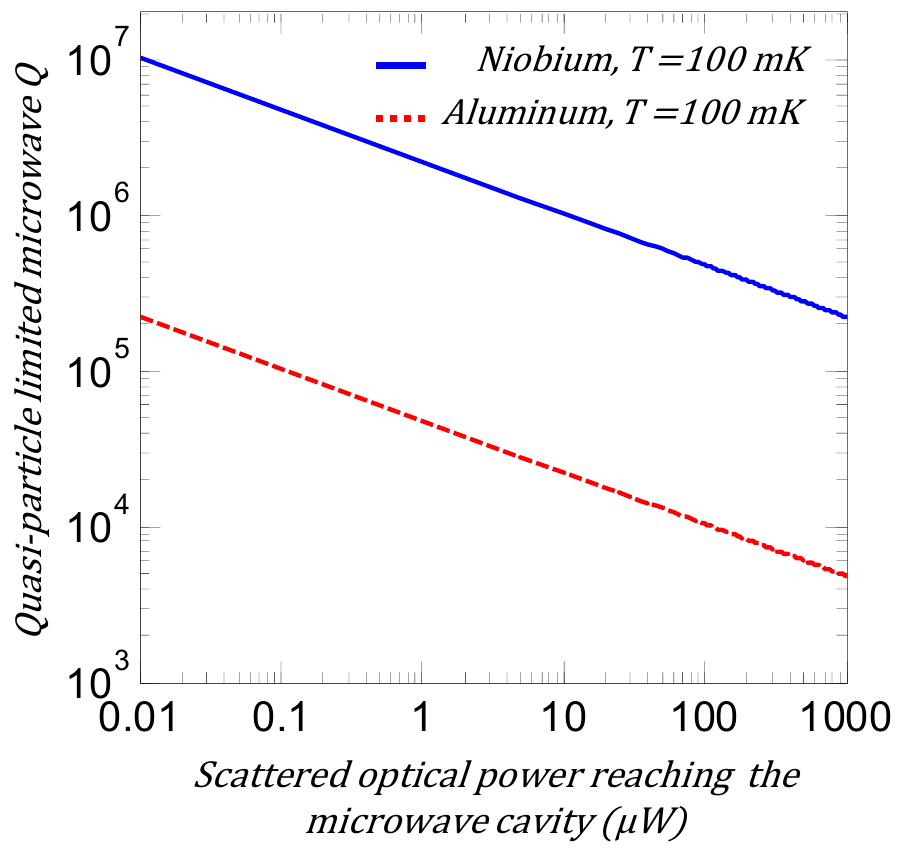}\caption{\label{fig:10}Variation of the microwave cavity $Q$ due to quasi-particles generated by optical photons scattered out from the optical resonator and reach the microwave cavity. This simulation considers a worst case scenario where all the scattered optical photons are absorbed by the microwave cavity.}
\end{figure}

\section{Discussion\label{sec:discuss}}
The results shown in Figs. 8 and 9 assume moderate values for the $g$ factor as well as the $Q$ of the optical resonator which is achievable with the present technology \cite{wang2017nanophotonic,moore2016efficient}. With further technology advances, achieving optical $Q > 10^6$ is quite feasible for LN resonators. This improvement in optical $Q$ together with increase in $g$ can dramatically reduce the pump power to sub microwatt scale. Having a low optical pump power allows designing a simpler filter for rejecting the pump photons and bandpass filtering the upconverted sideband which has one or few photons.

Important consideration is the effect of optical photons scattered from the optical resonator and reach the SC microwave resonator. These photons can generate quasi particles that can reduce the microwave $Q$ \cite{kaplan1976quasiparticle}. Assuming the exceedingly pessimistic scenario that all scattered optical pump photons are directed toward the microwave resonator and get absorbed there, we calculated the microwave $Q$ using a one-dimensional Rothwarth-Taylor model for quasiaparticle density \cite{Rothwarf1967, barends2011minimizing}. 

The results are summarized in Fig. 10, in the case of two SC materials - Aluminum and Niobium - at a temperature of 100 mK. As seen from this figure, even with scattered photons in the range of tens of microwatts, the $Q$ of the SC microwave resonator is above $10^4$. In reality, only fraction of scattered photons will reach the SC and be absorbed (some will reflect) ensuring that quasi particle limited microwave $Q$ stays well above $10^4$ for Aluminum SC, and well above $10^5$ for Niobium SC.

\section{Conclusion\label{sec:conclusion}}
We have presented an electro-optic based nanophotonic coupled-resonator with doublet resonances for efficient microwave-to-optical conversion of a SC microwave qubit. The frequency doublet spacing of the photonic coupled-resonator is matched to the microwave resonance of the qubit. The compactness of this design is shown to significantly increase the $g$ factor. The doublet spacing can be dynamically tuned allowing a perfect matching between the microwave resonance and the doublet spacing. We derived relation between the optical powers and the parameters of the optical and the microwave resonator as well as the $g$ factor for the optimal conversion process.

Our theoretical investigations show that with such coupled-resonator device implemented with LN nanophotonic resonators,  $g/2\pi$ in the range of $\sim 5-15$ kHz and beyond is achievable. While in our analysis a Z-cut LN was considered, the X-cut LN can be used as well, though the electrode placements needs to be modified. We show that with such $g$ values, and with moderate optical and the microwave $Q$ values, an optical pump power in the range of tens of microwatt is required for efficient microwave-to-optical conversion. Integrating this nanophotonic device with a SC qubit on the same platform is quite feasible, and promises a scalable coherent state transfer from SC quantum processors over an optical fiber, thus paving the way for networked quantum computing and communication.

\begin{acknowledgments}
We gratefully acknowledge the fruitful discussion with Zac Dutton, Leonardo Ranzani, Marcus Silva, Tom Ohki, and Borja Peropadre from Raytheon BBN, and Yaowen Hu from Harvard University on this manuscript. This document does not contain technology or technical data controlled under either the U.S. International Traffic in Arms Regulations or the U.S. Export Administration Regulations. Harvard team acknowledges partial support by NSF (grant No. ECCS-1609549) and ONR MURI  (Grant No. N00014-15-1-2761).
\end{acknowledgments}

%\bibliography{references}
%merlin.mbs apsrev4-1.bst 2010-07-25 4.21a (PWD, AO, DPC) hacked
%Control: key (0)
%Control: author (8) initials jnrlst
%Control: editor formatted (1) identically to author
%Control: production of article title (-1) disabled
%Control: page (0) single
%Control: year (1) truncated
%Control: production of eprint (0) enabled
%

\end{document}